\documentclass[%
 reprint,
 superscriptaddress,
amsmath,amssymb,
aps,
prx,
]{revtex4-2}

\usepackage[pdftex]{graphicx}
\usepackage{dcolumn}
\usepackage{bm}
\usepackage[pdftex]{hyperref}
\usepackage{color}
\usepackage{here}

\usepackage{here}

\begin{document}
\title{Feedback-driven quantum reservoir computing for time-series analysis}
\author{Kaito Kobayashi}
\affiliation{Department of Applied Physics, The University of Tokyo, 7-3-1 Hongo, Bunkyo-ku, Tokyo 113-8656, Japan}
\author{Keisuke Fujii}
\affiliation{Graduate School of Engineering Science, Osaka University, 1-3 Machikaneyama, Toyonaka, Osaka 560-8531, Japan}
\affiliation{Center for Quantum Information and Quantum Biology, Osaka University, 1-2 Machikaneyama, Toyonaka 560-0043, Japan}
\affiliation{RIKEN Center for Quantum Computing (RQC), Hirosawa 2-1, Wako, Saitama 351-0198, Japan}
\author{Naoki Yamamoto}
\affiliation{Department of Applied Physics and Physico-Informatics, Keio University, Hiyoshi 3-14-1, Kohoku-ku, Yokohama 223-8522, Japan}
\affiliation{Quantum Computing Center, Keio University, Hiyoshi 3-14-1, Kohoku-ku, Yokohama 223-8522, Japan}
\date{\today}

\begin{abstract}
  Quantum reservoir computing (QRC) is a highly promising computational paradigm that leverages quantum systems as a computational resource for nonlinear information processing. 
  While its application to time-series analysis is eagerly anticipated, prevailing approaches suffer from the collapse of the quantum state upon measurement, resulting in the erasure of temporal input memories. 
  Neither repeated initializations nor weak measurements offer a fundamental solution, as the former escalates the time complexity while the latter restricts the information extraction from the Hilbert space. 
  To address this issue, we propose the feedback-driven QRC framework.
  This methodology employs projective measurements on all qubits for unrestricted access to the quantum state, with the measurement outcomes subsequently fed back into the reservoir to restore the memory of prior inputs. 
  We demonstrate that our QRC successfully acquires the fading-memory property through the feedback connections, a critical element in time-series processing.
  Notably, analysis of measurement trajectories reveal three distinct phases depending on the feedback strength, with the memory performance maximized at the edge of chaos.
  We also evaluate the predictive capabilities of our QRC, demonstrating its suitability for forecasting signals originating from quantum spin systems. 
\end{abstract}

\maketitle

\section{INTRODUCTION}

In the modern information society, quantum machine learning is gaining heightened interest for its potential in advanced data processing~\cite{Wiebe:PRL:2012,Peruzzo:NatComm:2014,Biamonte:Nature:2017,McClean:NatCommun:2018,Cong:NatPhys:2019,Cerezo:NatRevPhys:2021,Endo:JPSJ:2021,Huang:science:2022-1,Huang:science:2022-2}. 
The cornerstone of the advantage lies in the use of the exponentially large Hilbert space, which allows for the accommodation of high-dimensional features of datasets and the emulation of a substantial number of computational neurons~\cite{Rebentrost:PRL:2014,Mitarai:PRA:2018,Dunjko:RepProgPhys:2018,Schuld:PRL:2019}. 
A quantum extension of the reservoir computing paradigm, known as quantum reservoir computing (QRC), also leverages this high dimensionality~\cite{Fujii:PRA:2017,Nakajima:PRA:2019}. 
Central to the philosophy of (general) reservoir computing is its distinctive architecture, where a fixed, randomly connected network of nodes--``reservoir''--processes input data by projecting it onto an internal feature space~\cite{Jaeger:GMD:2001,Jaeger:Science:2004,Verstraeten:NewralNetw:2007,Jaeger:NeuralNetw:2007,Tanaka:NeuralNetw:2019,Kobayashi:SciRep:2023,Kobayashi:NPSM:2023}. 
This strategy limits the training parameters only to the weights on the read-out part of the network, significantly reducing optimization costs and enabling high-speed, real-time temporal processing. 
Previous investigations have demonstrated that computational capabilities tend to be maximized when the reservoir part resides proximate to the boundary separating a stable and an unstable dynamical regimes, commonly referred to as the edge of chaos~\cite{Bertschinger:NeuralComp:2004,Boedecker:TheoryBiosci:2012,Snyder:PRE:2013,Carroll:Chaos:2020,Hochstetter:NatCommun:2021,Nishioka:SciAdv:2022}. 
Importantly, given the fixed nature of the internal parameters, the reservoir part can be realized either with artificial recurrent neural networks or with physical systems. 
In the QRC framework, a quantum system is harnessed as the physical embodiment of the reservoir part, termed a quantum reservoir. 
The intrinsic nonlinearity and high dimensionality of quantum systems perfectly align with the heuristic requirements for effective reservoir computing systems, which facilitate the extraction and transformation of intricate data features~\cite{Fujii:PRA:2017,Nakajima:PRA:2019,Tanaka:NeuralNetw:2019}.
Thus far, theoretical proposals of the QRC have yielded promising results in addressing temporal and classification tasks~\cite{Ghosh:npjqi:2019,Markovi:APL:2020,PRXQ:Bravo:2022,Sakurai:PRAppl:2022,Nokkala:SciRep:2023,Dudas:npj:2023,Hayashi:PRA:2023,Sakurai:arXiv:2024}, and experimental demonstrations have been reported on several quantum reservoir settings, including nuclear magnetic resonance (NMR) systems~\cite{arXiv:Negoro:2018} and superconducting qubits~\cite{Chen:PRA:2020}.
Furthermore, since the information processing capabilities of the QRC are fundamentally linked to the nature of the quantum reservoirs~\cite{Pena:PRL:2021,Xia:FrontPhys:2022}, it can be employed as a tool to investigate the quantum system itself, a concept recently formalized as quantum reservoir probing~\cite{Kobayashi:arXiv:2023,Kobayashi:arXiv:2024}.

Particularly focusing on temporal information processing, the fading-memory property emerges as a crucial aspect of the reservoir computing framework. 
This property dictates that when the reservoir projects input data onto the feature space, the temporal dependencies between different inputs are faithfully mapped within that internal space, and the influences of earlier inputs, being less significant than more recent ones, gradually diminish~\cite{Jaeger:GMD:2001,Jaeger:Science:2004}. 
Herein lies a significant challenge for the QRC: the inherent conflict between information extraction and memory preservation within the quantum reservoir. 
Conventional QRC architectures rely on quantum measurements to extract information from the quantum reservoir. 
However, these measurements inevitably induce measurement back-action, leading to the destruction of the quantum state of the reservoir. 
As a result, the memories of previous inputs encoded within the quantum reservoir are completely erased. 
This disturbance compromises the preservation of short-term memory and thus hinders the ability to process temporal information due to the absence of the crucial fading-memory property.

\begin{figure}[t!]
  \centering
  \includegraphics[width=\hsize]{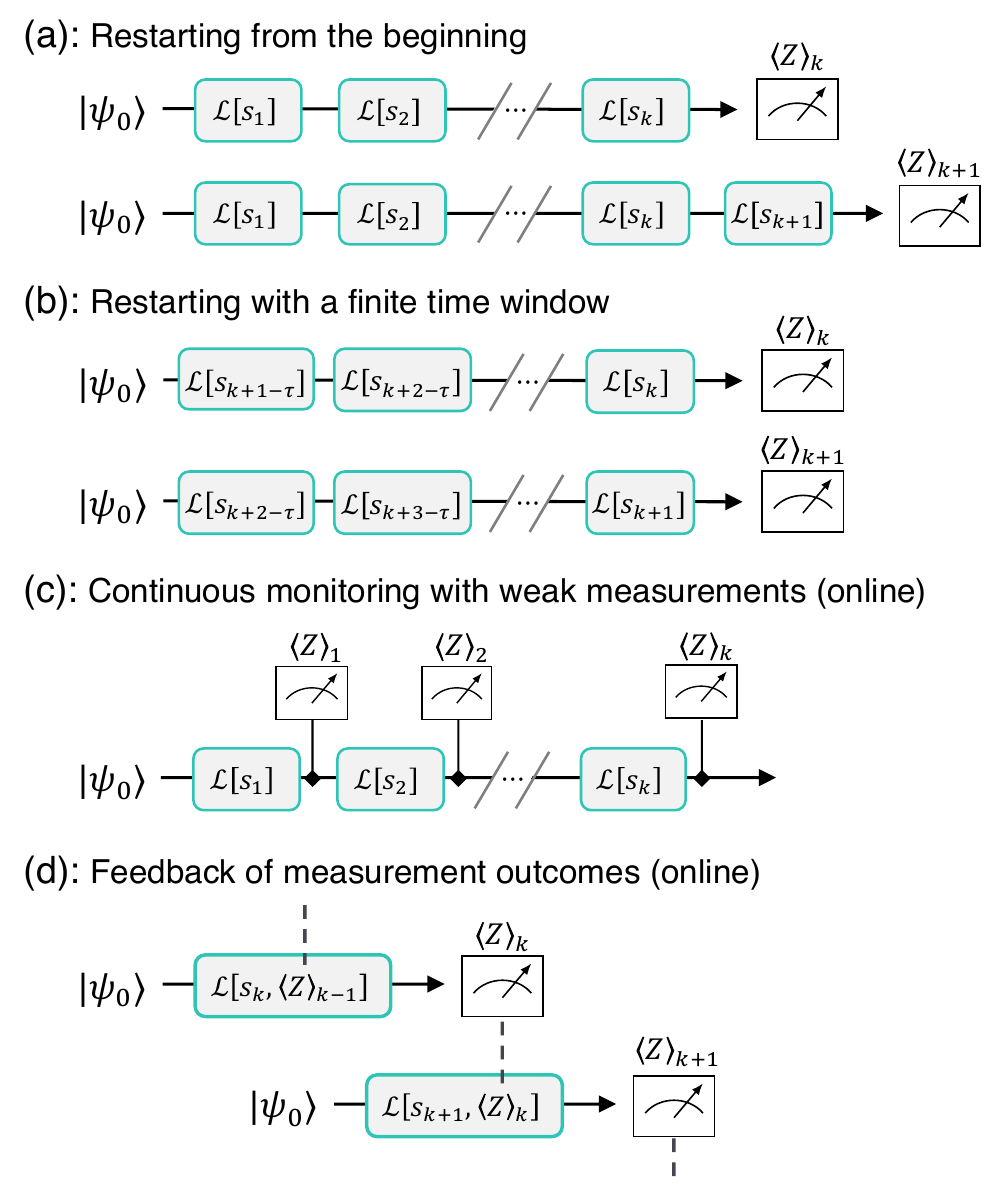}
  \caption{
    Architectures of QRC with an input sequence \(\{s_k\}\) and an initial state \(|\psi_0\rangle\). 
    \(\mathcal{L}\) symbolizes a generalized operation for a single input cycle parameterized by external inputs, the specific form of which depends on each realization. 
    (a) Quadratic restarting protocol where the entire input signal is provided at every cycles to reinitialize the reservoir after the measurements. 
    (b) Linear restarting protocol where the last \(\tau\) inputs are repeated, assuming less effects from further previous inputs. 
    (c) Online protocol that continuously monitors the time evolution using weak measurements or projective measurements only on limited part of the system. 
    (d) Feedback protocol where the measurement result at the \((k-1)\)-th cycle determines the operation \(\mathcal{L}\) at the \(k\)-th cycle. 
  }\label{fig1}
\end{figure}

To address this monitoring issue, a common strategy involves repeated initialization after each measurement. 
The entire process is executed from the beginning for every individual input, thereby effectively neglecting the effects of measurement back-actions [Fig.~\ref{fig1}(a)]~\cite{Chen:PRA:2020,Suzuki:SciRep:2022,Fry:SciRep:2023,Kubota:PRA:2023}. 
However, this approach naturally necessitates \(O(L^2)\) input operations to address a time-series of length \(L\), leading to a computationally expensive quadratic growth in execution time. 
A potential reduction in temporal complexity could be achieved under the premise of the fading-memory property. 
As the reservoir gradually loses information about distant past inputs, the simplification of the process by limiting input repetitions to recent cycles is expected to have minimal impact on reservoir performance [Fig.~\ref{fig1}(b)]. 
By initiating the dynamics for the input cycle \(k\) from the cycle \(k-\tau+1\) rather than the cycle \(1\), the total number of input repetitions reduces to \(\tau L\), yielding a time complexity of \(O(L)\)~\cite{Vcindrak:PRR:2024}. 
Nevertheless, despite this improvement, the requirement for repeating gate operations to input identical data still significantly compromises the real-time processing capabilities, a key advantage of the reservoir computing paradigm.

A promising approach seeks to balance short-term memory preservation with real-time functionality, by continuously monitoring quantum dynamics online while partially retaining the reservoir's quantum state at each measurement [Fig.~\ref{fig1}(c)]. 
By employing weak measurements or projective measurements on a limited subset of qubits, the QRC generates output online while mitigating the erasure of past input memory~\cite{Takaki:PRA:2021,Mujal:npjqi:2023,Yasuda:arXiv:2023}. 
Continuous variable quantum optical systems offer efficient platforms for this architecture~\cite{Nokkala:CommPhys:2021,Garcia-Beni:PRAppl:2023,Nokkala:MachLearnSciTechnol:2024}. 
However, these approaches only permit limited information extraction, as partial measurements do not provide comprehensive access to the quantum state of the reservoir. 
This constraint hinders the full utilization of the Hilbert space in which the quantum reservoir resides. 
Given that the exponential scaling of dimensionality with system size is a fundamental strength of the QRC, this may constitute a drawback when compared to alternative configurations of comparable system size, although, the extent to which it negatively affects performance would depend on the specific implementations. 
Consequently, the development of a simple and versatile online QRC protocol that fully leverages the Hilbert space of the quantum reservoir remains an urgent area of research. 

In this paper, we propose a feedback-driven online QRC protocol that achieves complete access to the quantum state through projective measurements on all qubits [Fig.~\ref{fig1}(d)]. 
While these measurements destroy short-term memory within the reservoir's state, information of prior inputs is effectively revived by feedback connections that incorporate the measurement outcomes back into the quantum reservoir~\cite{Pfeffer:PRR:2022,Pfeffer:PRR:2023,Wudarski:arXiv:2023}. 
By evaluating the short-term memory performance, we demonstrate that our QRC framework successfully acquire the fading-memory property through the feedback, showcasing its applicability to time-series analysis. 
The proposed QRC system is highly tunable, as the strength of input and feedback, which strongly relates to information embedding and memory preservation, can be controlled externally. 
Notably, analysis of the measurement trajectories reveals three distinct phases depending on the feedback strength. 
The memory performance peaks at the phase boundary between the stable and the unstable phase, i.e., the edge of chaos. 
We further evaluate the predictive performance of our QRC for various time-series signals. 
Benchmark analysis against a conventional classical reservoir computing framework reveals that our QRC is highly suitable for forecasting signals derived from quantum spin systems. 
Our findings highlight the substantial potential of the QRC in the burgeoning quantum era as a potent instrument for time-series analysis. 

The paper is organized as follows. 
In Sec.~II, we introduce the framework of our feedback-driven QRC scheme. 
In Sec.~III, we evaluate the short-term memory performance of the QRC. 
An in-depth analysis of the measurement trajectories, including the identification of the edge of chaos, is presented therein. 
In Sec.~IV, we examine the prediction performance for the classical and quantum time-series signals. 
Section V is devoted to discussion and conclusion.

\section{METHODS}
\subsection{Architecture of the feedback-driven QRC}\label{sec2a}

\begin{figure}[t!]
  \centering
  \includegraphics[width=\hsize]{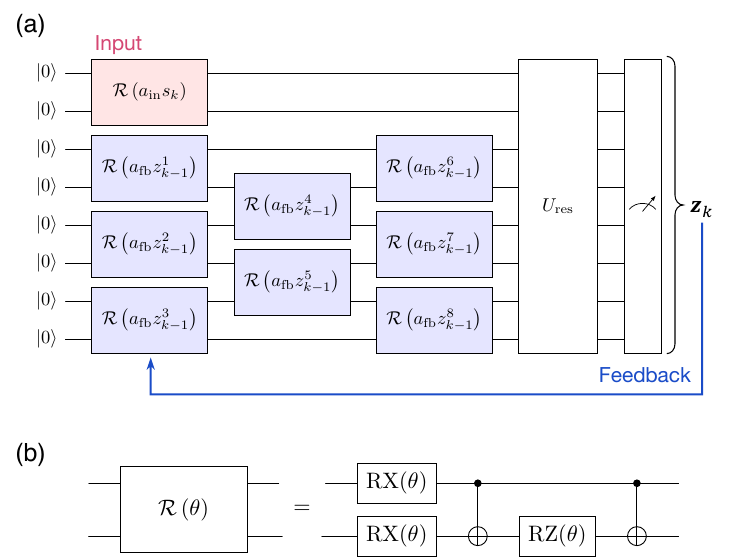}
  \caption{
    (a) Circuit diagram of the feedback-driven QRC architecture at the \(k\)-th cycle. 
    The red and blue background colors represent the gates for the input [step (i)] and feedback [step (ii)] parts, respectively. 
    The input value \(s_k\) is provided to qubits \(1\) and \(2\) via \(\mathcal{R}_{1,2}(a_{\mathrm{in}}s_k)\) with the input weight \(a_{\mathrm{in}}\). 
    The feedback of the measurement result \(\bm{z}_{k-1}\) is given to qubits \(3\) to \(8\) via \(\mathcal{R}_{i,j}(a_{\mathrm{fb}}z_{k-1}^\alpha)\). 
    Subsequently, the entanglement among all qubits is introduced by the reservoir gate \(U_{\mathrm{res}}\) [step (iii)], and all qubits are then measured in the projective manner at the end of the circuit [step (iv)]. 
    (b) Circuit representation of the two-qubit operator \(\mathcal{R}\) defined in Eq.~(\ref{eq1}).
  }
  \label{fig2}
\end{figure}

Figure \ref{fig2}(a) illustrates the architecture of our feedback-driven QRC system. 
We consider an \(N\)-qubit system initialized in the state \(|\psi\rangle_0=|0\rangle^{\otimes N}\); in our numerical experiment, we specifically set \(N=8\).  
Our QRC protocol consists of four steps: (i) input operation, (ii) feedback of previous measurement results, (iii) reservoir dynamics, and (iv) projective measurement on all qubits.

In step (i) of the \(k\)-th cycle, the input value \(s_k\) is injected into the reservoir via the application of a two-qubit gate \(\mathcal{R}_{1,2}(a_{\mathrm{in}}s_k)\) on qubits \(1\) and \(2\), where \(a_{\mathrm{in}}\) denotes an input scaling weight. 
The operator \(\mathcal{R}_{i,j}\) is defined as
\begin{equation}
  \mathcal{R}_{i,j}(\theta)=\mathrm{CX}_{ij}\mathrm{RZ}_j(\theta)\mathrm{CX}_{ij}\mathrm{RX}_i(\theta)\mathrm{RX}_j(\theta), \label{eq1}
\end{equation}
which is commonly employed for embedding information in the QRC protocols~\cite{Suzuki:SciRep:2022,Fry:SciRep:2023,Kubota:PRA:2023}. 
Here, \(\mathrm{CX}_{ij}\) denotes the \textrm{CNOT} gate acting on control qubit \(i\) and target qubit \(j\), while \(\mathrm{RZ}(\theta)\) and \(\mathrm{RX}(\theta)\) represent the rotation gate around the Pauli \(Z\) and \(X\) axis with angle \(\theta\), respectively. 
The circuit diagram for \(\mathcal{R}_{i,j}(\theta)\) is depicted in Fig.~\ref{fig2}(b). 

For clarity, we next describe the step (iv) of the (\(k-1\))-th cycle. 
Here, Pauli \(Z\) measurements are executed on all qubits, which yield the measurement result vector \(\bm{z}_{k-1}=[\langle Z_1\rangle_{k-1},\langle Z_2\rangle_{k-1},\dots,\langle Z_{N}\rangle_{k-1}]^\top\) (where \(Z_i\) denotes the Pauli \(Z\) for the qubit \(i\)). 
Returning to step (ii) of the \(k\)-th cycle, the vector \(\bm{z}_{k-1}\) is provided back into the reservoir through the feedback connection. 
The operator \(\mathcal{R}_{i,j}(a_{\mathrm{fb}}z_{k-1}^\alpha)\) is employed for feedback with a weight \(a_{\mathrm{fb}}\), which encodes all components of \(\bm{z}_{k-1}\) according to the configuration shown in Fig.~\ref{fig2}(a). 
Although the projective measurement strongly perturbs the quantum state, the feedback connections operate to systematically restore the memory of the provided input history. 

Subsequent to the input and feedback processes, a fixed eight-qubit gate \(U_{\mathrm{res}}\) is applied in step (iii) to generate entanglement among all qubits. 
This gate serves as the reservoir component of our QRC architecture, which simultaneously transforms the input and feedback information in a non-linear manner. 
We employ a Haar random unitary matrix for \(U_{\mathrm{res}}\) in the following calculations: refer to Appendix \ref{AppendixA} for a hardware-efficient implementation of \(U_{\mathrm{res}}\). 
Finally, step (iv) concludes the \(k\)-th cycle with projective measurements on all qubits under the assumption of negligible statistical fluctuations. 
The measurement result \(\bm{z}_k\) is further cycled back to the reservoir via the feedback gate for processing the subsequent \((k+1)\)-th input. 
We note that additional feedback layers can be integrated to capture long-term correlations in data, as illustrated in Appendix \ref{AppendixB}.

It is worth highlighting a key advantage of our architecture in terms of time complexity, which arises from the intrinsic independence of quantum circuits over different cycles. 
In our QRC, the quantum state at the \(k\)-th cycle depends exclusively on the current input value \(s_k\) and the preceding measurement outcome \(\bm{z}_{k-1}\), explicitly represented as \(|\psi(s_k,\bm{z}_{k-1})\rangle\). 
Even after the measurement, this state can be efficiently reconstructed by applying a limited set of gates, leveraging  \(\bm{z}_{k-1}\) stored in external memory; there is no need to re-execute the \((k-1)\)-th and prior cycles when addressing the \(k\)-th input. 
Consequently, our protocol offers remarkable operational flexibility, enabling features such as starting from an intermediate cycle or running different cycles on different devices. 
In addition, as the circuit terminates with measurements after each individual cycle, a coherence time sufficient for the execution of just one cycle is adequate for our protocol. 
This aspect stands in stark contrast to other implementations depicted in Figs.~\ref{fig1}(a) and \ref{fig1}(c), which, in order to process temporal data up to the \(L\)-th input cycle, necessitate \(L\) repetitions of the reservoir operations to construct the history-dependent quantum state \(|\psi(s_k,s_{k-1},s_{k-2}\dots)\rangle\). 
Assuming a single cycle duration of \(T\), a limitation on the input data length \(L\) is thus imposed to ensure that the operation time \(LT\) remains within the bound of the coherence time. 
In the protocol shown in Fig.~\ref{fig1}(b), the required coherence time is reduced to \(\tau T\) by limiting the number of input repetitions at each cycle to \(\tau\); however, the need for a large \(\tau\) to enhance performance is still in conflict with the temporal constraint imposed by the coherence time. 
In light of this comparison, our feedback-driven protocol necessitates the most minimal coherence time among these approaches, making it particularly well-suited for hardware implementations.

\subsection{Learning procedure}

In the QRC protocol, the final output \(y_k\) at the \(k\)-th cycle is obtained through a linear transformation of the measurement result \(\bm{z}_k\). 
Here, a specific task for the QRC is formulated by setting a target output \(\{\bar{y}_k\}\) based on the input sequence \(\{s_k\}\). 
Our objective in the output part is to construct a regression model from the collection of measurement outcomes that accurately approximates the target output over various cycles. 
As a prerequisite, we define \(\bm{z}_0\), required for the first cycle (\(k=1\)), as a random vector with each element sampled uniformly from the interval \([0,1]\). 
The influence of this initial condition is disregarded by discarding the initial \(l_{\mathrm{w}}\) cycles. 

The linear regression model is optimized for a training dataset containing \(l_{\mathrm{tr}}\) inputs, whose performance is subsequently evaluated for unseen data in the testing phase for \(l_{\mathrm{ts}}\) inputs. 
We define an \([l_{\mathrm{tr}}\times(N+1)]\)-dimensional training matrix \({X}_{\mathrm{tr}}\) from the measurement results as 
\begin{equation}
  {X}_{\mathrm{tr}} = [{\bm{z}'}_{l_{\mathrm{w}}+1}, {\bm{z}'}_{l_{\mathrm{w}}+2} \ldots {\bm{z}'}_{l_{\mathrm{w}}+l_{\mathrm{tr}}}]^\top,\label{eq2}
\end{equation}
where \(\bm{z}_k'=\left[\bm{z}_k^\top,1\right]^\top\) including an additional constant term. 
In the training phase, the weight vector \(\bm{w}\) is optimized so that the output \(\bm{y}_{\mathrm{tr}}={X}_{\mathrm{tr}}\bm{w}\) closely approximates the target value \(\bar{\bm{y}}_{\mathrm{tr}}\). 
Utilizing the Moore-Penrose pseudo-inverse matrix, the optimal \(\bm{w}\) is given by 
\begin{equation}
  \bm{w} = (X^\top_{\mathrm{tr}} X_{\mathrm{tr}})^{-1}X^\top_{\mathrm{tr}}\bar{\bm{y}}_{\mathrm{tr}}.
\end{equation}

Following the training phase, we construct an \([l_{\mathrm{ts}}\times(N+1)]\)-dimensional matrix \({X}_{\mathrm{ts}}\) for testing in a manner analogous to Eq.~(\ref{eq2}), using the measurement outcomes from the (\(l_{\mathrm{w}}+l_{\mathrm{tr}}+1\))-th cycle to the (\(l_{\mathrm{w}}+l_{\mathrm{tr}}+l_{\mathrm{ts}}\))-th cycle. 
The output is then calculated as \(\bm{y}_{\mathrm{ts}}={X}_{\mathrm{ts}}\bm{w}\) using the trained weight. 
The similarity between the output \(\bm{y}_{\mathrm{ts}}\) and the target output \(\bar{\bm{y}}_{\mathrm{ts}}\) in the testing phase signifies the reservoir performance. 
As a quantitative metric, we utilize the determination coefficient
\begin{equation}
  R^2 = \frac{\mathrm{cov}^2(\bar{\bm{y}}_{\mathrm{ts}},\bm{y}_{\mathrm{ts}})}{\sigma^2(\bar{\bm{y}}_{\mathrm{ts}})\sigma^2(\bm{y}_{\mathrm{ts}})},
\end{equation} 
where \(\mathrm{cov}\) and \(\sigma^2\) denote covariance and variance, respectively. 
\(R^2\) approaches \(1\) when the output \(\bm{y}_{\mathrm{ts}}\) closely matches the target \(\bar{\bm{y}}_{\mathrm{ts}}\); conversely, \(R^2\) approaching \(0\) indicates a lack of correlation. 
We also employ the Normalized Mean Squared Error (NMSE):
\begin{equation}
    \label{eq5}
  \mathrm{NMSE} = \frac{||\bar{\bm{y}}_{\mathrm{ts}}-\bm{y}_{\mathrm{ts}}||^2}{||\bar{\bm{y}}_{\mathrm{ts}}||^2},
\end{equation}
where $|| \bullet ||$ denotes the Euclidean norm. 
The NMSE can assume any real value, and a low value of NMSE indicates a small deviation between the output and target. 
In our setup, we evaluate the reservoir performance with \(l_{\mathrm{w}}=500\), \(l_{\mathrm{tr}}=2000\), and \(l_{\mathrm{ts}}=2000\).

\section{Fading-memory property with feedback connections}\label{sec3}
\subsection{Short-term memory capacity}

To evaluate the effectiveness of feedback connections for the preservation of input memories, we investigate the short-term memory task. 
This task involves processing an input sequence \(\{s_k\}\) randomly sampled from a uniform distribution: \(s_k\in[0,1]\). 
The goal is to reproduce the input value from \(d\)-cycle earlier, formulated as \(\bar{y}_k=s_{k-d}\). 
The performance is measured using the determination coefficient \(R^2_d\) for varying delays \(d\), and the total capacity \(C_{\Sigma}\) is defined by summing up these values
\begin{equation}
  C_{\Sigma}\equiv\sum_{d=0}^{d_{\mathrm{max}}}R^2_d, 
\end{equation}
where we set \(d_{\mathrm{max}}=25\). 
Both \(R^2_d\) and \(C_{\Sigma}\) are averaged over \(128\) samples with respect to the random matrix \(U_{\mathrm{res}}\). 

\begin{figure}[t!]
  \centering
  \includegraphics[width=\hsize]{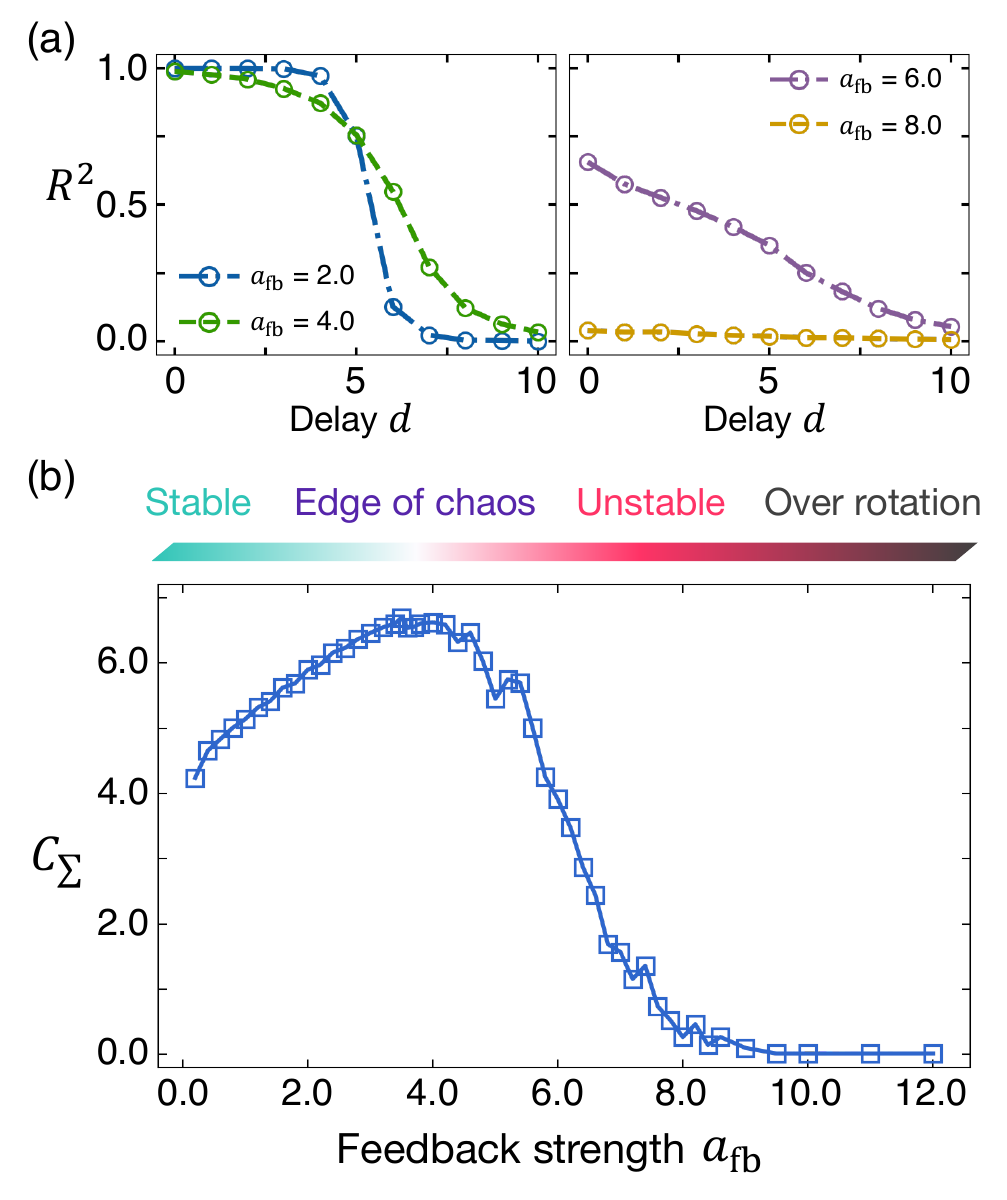}
  \caption{  
    (a) Reservoir performance for the short-term memory task \(R^2_{d}\) plotted as a function of the delay \(d\) with \(a_{\mathrm{fb}}=2.0\) (blue), \(a_{\mathrm{fb}}=4.0\) (green), \(a_{\mathrm{fb}}=6.0\) (purple), and \(a_{\mathrm{fb}}=8.0\) (yellow). 
    (b) The dependency of the total capacity \(C_{\Sigma}\) on the feedback strength \(a_{\mathrm{fb}}\). 
    Both \(R^2_{d}\) and \(C_{\Sigma}\) are averaged over \(128\) samples with respect to \(U_{\mathrm{res}}\). 
  }
  \label{fig3}
\end{figure}

Figure \ref{fig3}(a) exhibits the memory performance \(R^2_d\) for varying feedback weights \(a_{\mathrm{fb}}\) with the input weight fixed at \(a_{\mathrm{in}}=0.001\). 
At the feedback strength of \(a_{\mathrm{fb}}=2.0\), \(R^2_d\) becomes close to \(1\) for delays up to \(d=4\), followed by a gradual decline for longer delays. 
Considering that the quantum state of the \(k\)-th cycle is determined only by \(s_k\) and \(\bm{z}_{k-1}\), this robust memory performance indicates that the input histories of \(\{s_k\}\) are adequately accumulated within the measurement results \(\bm{z}_{k-1}\). 
The quantum reservoir effectively extracts key features of the provided dataset through the read-out measurements, and these features are succeeded over different cycles via the feedback connections. 
In this manner, our feedback-driven QRC overcomes the destruction of the quantum state caused by the measurement back-action, thereby acquiring the fading-memory property. 

When the feedback is intensified to \(a_{\mathrm{fb}}=4.0\), \(R^2_d\) slightly diminishes for smaller delays (\(d\leq 4\)), as the influence of inputs becomes relatively subdued compared to the feedback effects. 
Conversely, for longer delays \(d\geq 5\), \(R^2_d\) shows an improvement characterized by a long-tailed structure as a result of the amplified feedback. 
The balance between the encoding and retention of the supplied information is thus tuned by \(a_{\mathrm{fb}}\). 
Notably, excessively strong feedback destabilizes the system, evidenced by a decrease in \(R^2_d\) at the feedback strength of \(a_{\mathrm{fb}}=6.0\). 
Furthermore, at the even higher feedback strength (\(a_{\mathrm{fb}}=8.0\)), the rotation angle \(a_{\mathrm{fb}}z^i_{k-1}\) in Eq.~(\ref{eq1}) exceeds \(2\pi\), causing a malfunction of the feedback gates and resulting in \(R^2_d\simeq0\) for all \(d\). 

Figure \ref{fig3}(b) depicts the total capacity \(C_{\Sigma}\) as a function of the feedback strength \(a_{\mathrm{fb}}\), revealing three distinct phases: a stable phase for \(0<a_{\mathrm{fb}}\lesssim 4.0\), an unstable phase for \(4.0\lesssim a_{\mathrm{fb}}\lesssim 8.0\), and an over-rotation phase for \(8.0\lesssim a_{\mathrm{fb}}\). 
In the stable phase, \(C_{\Sigma}\) shows an increasing trend with \(a_{\mathrm{fb}}\), which indicates that the stronger feedback more effectively revives older information. 
This phase represents a well-balanced state where the feedback mechanism enhances memory retention capabilities without excessively diminishing the influence of the new input. 
In contrast, the unstable phase is characterized by a decline in \(C_{\Sigma}\) as feedback intensifies, due to the system being predominantly governed by feedback rather than input. 
Such feedback dominance compromises the reservoir's functionality in processing and preserving information effectively. 
The over-rotation phase is marked by a complete collapse of \(C_{\Sigma}\) to zero, triggered by an excessively large \(a_{\mathrm{fb}}\) leading to the over-rotation upon the feedback. 
Consequently, optimal memory performance is achieved with feedback strength just below the threshold for the onset of unstable behavior~\cite{Pena:PRL:2021}. 
This point, in close proximity to the boundary between the stable and unstable phases, corresponds to the edge of chaos in this system. 

\subsection{Measurement trajectories}

\begin{figure}[t!]
  \centering
  \includegraphics[width=\hsize]{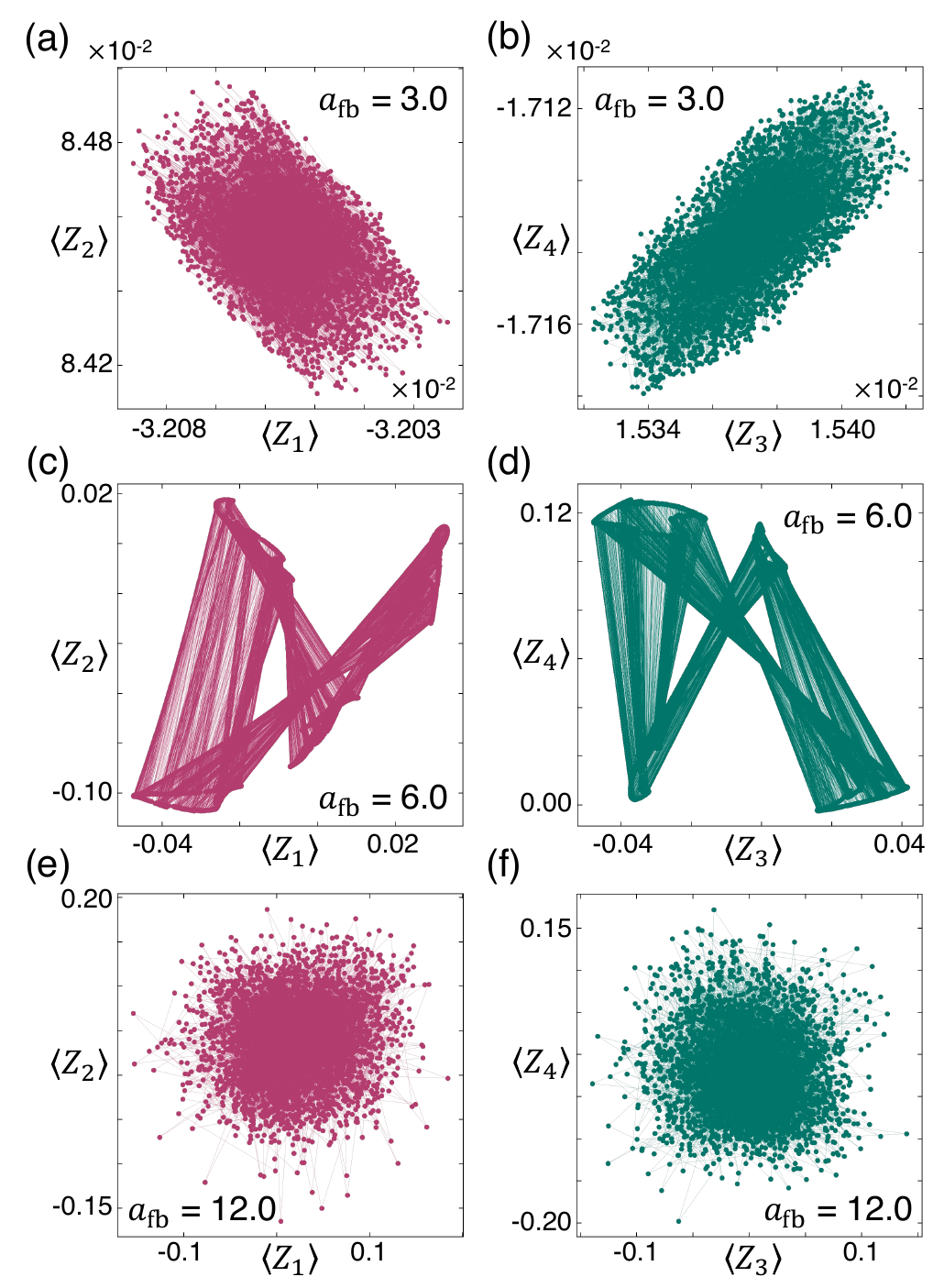}
  \caption{
    Trajectories of the measurement results in (a), (c), (e) \(\langle Z_1\rangle\)-\(\langle Z_2\rangle\) plane and (b), (d), (f) \(\langle Z_3\rangle\)-\(\langle Z_4\rangle\) plane for a typical instance of \(U_{\mathrm{res}}\). 
    Each point represents \(\left[\langle Z_i\rangle_k,\langle Z_{i+1}\rangle_k\right]\) at cycle \(k\), and each line connect \(\left[\langle Z_i\rangle_k,\langle Z_{i+1}\rangle_k\right]\) and \(\left[\langle Z_i\rangle_{k+1},\langle Z_{i+1}\rangle_{k+1}\right]\). 
    We employ \(a_{\mathrm{fb}}=3.0\) for (a) and (b), \(a_{\mathrm{fb}}=6.0\) for (c) and (d), and \(a_{\mathrm{fb}}=12.0\) for (e) and (f). 
  }
  \label{fig4}
\end{figure}

Since our QRC is driven by the feedback of the measurement outcomes and generates outputs as linear transformations thereof, an analysis of the measurement trajectories offers a deeper understanding into the physical origin underlying the three identified phases. 
Figure \ref{fig4} displays the measurement results \(\left[\langle Z_1\rangle_k,\langle Z_2\rangle_k\right]\) and \(\left[\langle Z_3\rangle_k,\langle Z_4\rangle_k\right]\) at each cycle \(k\) (similar trajectories are observed for \(\langle Z_5\rangle_k\) to \(\langle Z_8\rangle_k\)). 
For \(a_{\mathrm{fb}}=3.0\) in the stable phase, the dynamics of the quantum reservoir remain confined within a limited region of the feature space [Figs.~\ref{fig4}(a) and \ref{fig4}(b)]. 
This behavior is attributable to the interplay between the feedback, acting to restore the pre-measurement state, and the input, introducing perturbations. 
The attractor-shaped trajectory within the feature space thus encapsulates the combined influences of both input and feedback. 
This encoding mechanism provides an understanding of the substantial preservation of short-term memories within the quantum reservoir in the stable phase. 

However, the attractor-shaped dynamics collapses with further intensified feedback. 
Figures \ref{fig4}(c) and \ref{fig4}(d) illustrate typical dynamics in the unstable phase at \(a_{\mathrm{fb}}=6.0\). 
Here, the dominant feedback causes the system to cycle through multiple stable states in the feature space, where the trajectory converges into a limit cycle. 
This undermines the influence of the input on the dynamics and, consequently, diminishes the system's responsiveness to new inputs, leading to a degradation in memory performance to \(C_{\Sigma}\simeq 0\). 
Importantly, the formation of limit cycles depends on the specific realization of \(U_{\mathrm{res}}\). 
For a particular \(U_{\mathrm{res}}\) where the dynamics avoids entrapment in a limit cycle, attractor-like trajectories resembling those in Figs.~\ref{fig4}(a) and \ref{fig4}(b) still emerge with a nonzero \(C_{\Sigma}\) even at \(a_{\mathrm{fb}}=6.0\). 
Nonetheless, as the feedback strength intensifies, the system exhibits an increased tendency to get caught in the limit cycle at a variety of \(U_{\mathrm{res}}\) due to augmented instability. 
This results in a monotonic decrease in the capacity \(C_{\Sigma}\) averaged over multiple instances of \(U_{\mathrm{res}}\) in the unstable phase. 
In the over-rotation phase, the trajectory degrades into a random dynamics, with neither input nor feedback having a discernible influence [Figs.~\ref{fig4}(e) and \ref{fig4}(f)]. 

As these findings indicate, the dynamics of the QRC, and consequently its overall computational capabilities, are determined by the balance between the input and feedback operations. 
It is important to note that, in the presence of statistical uncertainty in measurements, the performance is similarly governed by the interplay among the input, feedback, and statistical fluctuations. 
While such fluctuations can perturb the dynamics of the QRC to a certain degree, their effects can be mitigated by flexibly adjusting the input and feedback strengths. 
Therefore, the overall behavior of our QRC system remains largely unchanged even when measurement fluctuations are considered; see Appendix \ref{AppendixC} for a more detailed analysis.

\section{Prediction of classical and quantum time-series signals}

\subsection{Dynamics of classical and quantum systems}

Having demonstrated the preservation of the short-term memory facilitated by the feedback connections, we here evaluate the predictive capability of our QRC protocol for both classical and quantum time-series signals. 
This task is formulated by considering an input sequence  \(\{s_k\}\) normalized within the interval \([0,1]\), extracted from a specific dynamical system via time discretization.
The aim of this task is to generate predictions \(\tau_f\) cycles into the future, formally represented as \(\bar{y}_k=s_{k+\tau_f}\). 
For performance evaluation of the QRC in predicting classical dynamics, we utilize the Mackey-Glass (MG) delay differential equation~\cite{Jaeger:Science:2004,Fujii:PRA:2017,Chen:NatComm:2022,Dudas:npj:2023}: 
\begin{equation}
  \dot{x}(t) = \frac{\alpha x(t-t_d)}{1+x(t-t_d)^\beta}-\gamma x(t).\label{defMG}
\end{equation}
With the parameters \(\alpha=0.2\), \(\beta=10\), and \(\gamma=0.1\), the system manifests a chaotic attractor for \(t_d>16.8\); our analysis centers on the MG dynamics with \(t_d=17\). 
In formulating the input time-series signals, a single time-evolution step in the MG dynamics from \(x(t)\) to \(x(t+1)\) is represented by a single input cycle from \(k\) to \(k + 1\) in the sequence \(\{s_k\}\). 
In Fig.~\ref{fig5}(a), the time-series is visually illustrated in the \(s_k\)-\(s_{k+10}\) plane, along with an explicit representation as a function of time step \(k\).

Additionally, to assess the efficacy in forecasting dynamics of quantum origin, we examine the spin dynamics of a 1D quantum Ising chain comprising five spins. 
The Hamiltonian is given by 
\begin{equation}
  \mathcal{H} = -J\sum_{i}\sigma^z_i\sigma^z_{i+1} + h_x \sum_{i} \sigma^x_i + h_z \sum_{i} \sigma^z_i.\label{defHdy}
\end{equation}
Here, \({\sigma}_i^x\) and \({\sigma}_i^z\) denote the \(x\) and \(z\) Pauli matrices at the site \(i\), while \(h_x\) and \(h_z\) signify the transverse and longitudinal magnetic fields, respectively. 
\(J>0\) corresponds to the strength of the ferromagnetic nearest-neighbor Ising interaction, normalized to \(J=1\) for our energy scale. 
Our focus lies in predicting the expectation value of the central spin's \(z\)-component, \(\langle\sigma^z_3(t)\rangle\), where the quantum system is initialized in the all up state and evolving under the Hamiltonian in Eq.~(\ref{defHdy}).
In the absence of a longitudinal field (\(h_z = 0\)), the spin system can be mapped to a free fermion system via the Jordan-Wigner transformation, leading to integrable dynamics~\cite{Sachdev:Cambridge:2011}.
Conversely, at \((h_x , h_z)= (1.05, -0.5)\), the system exhibits nonintegrable dynamics characterized by the chaotic spectral statistics~\cite{Banuls:PRL:2011}. 
The time-series \(\{s_k\}\) is constructed by mapping the dynamics of the quantum spin from \(t\) to \(t+0.05\) to a single step from \(k\) to \(k+1\); visual representations of these inputs are provided in Fig.~\ref{fig5}(a). 

\subsection{Predictive performance of the QRC}\label{sec4b}

\begin{figure}[t!]
  \centering
  \includegraphics[width=\hsize]{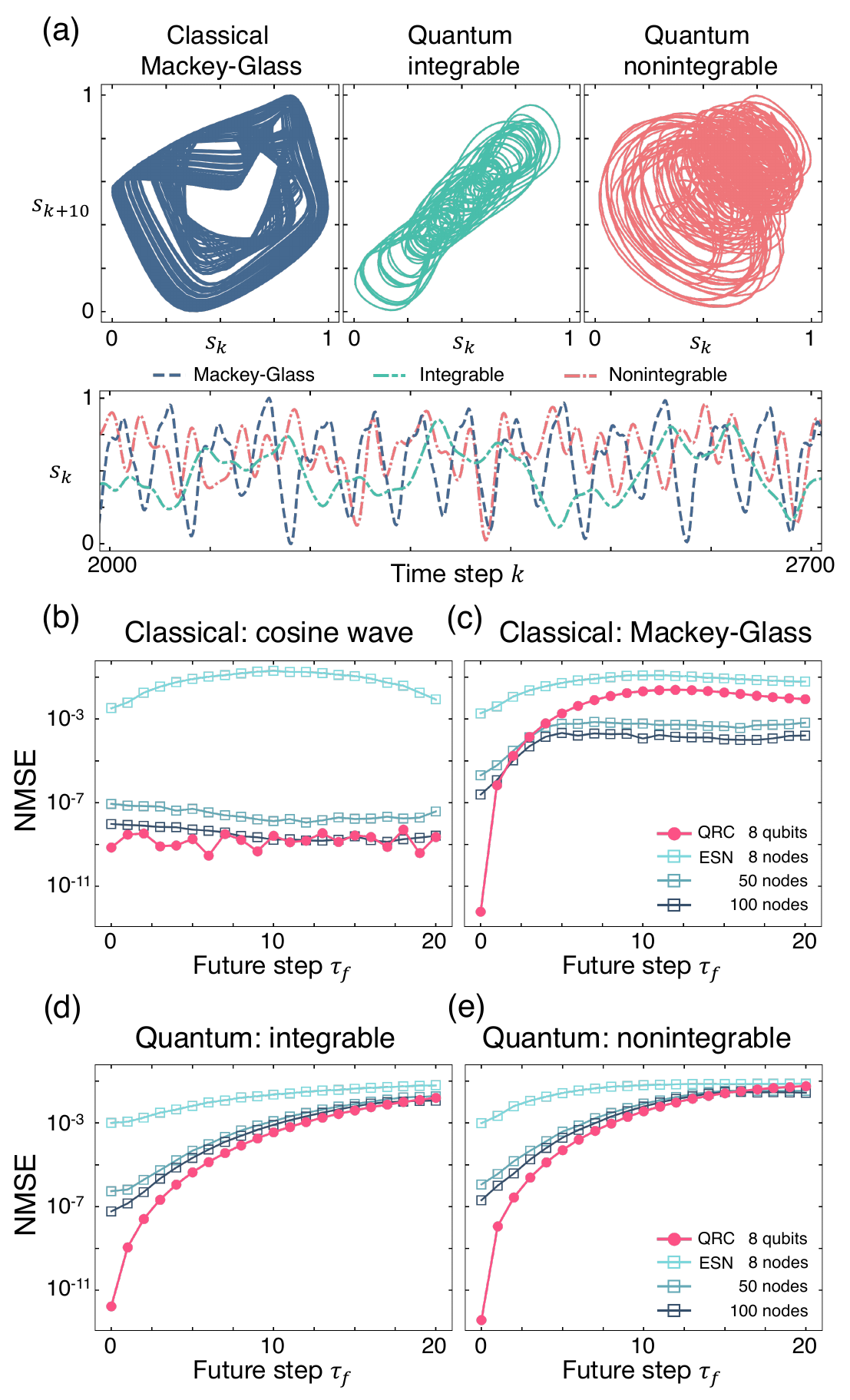}
  \caption{
    (a) Visualization of the time-series input signal \(\{s_k\}\) derived from classical MG dynamics, integrable quantum spin dynamics at \((h_x , h_z)= (1.0, 0.0)\), and nonintegrable quantum spin dynamics at \((h_x , h_z)= (1.05, -0.5)\). Each \(\{s_k\}\) is depicted in the \(s_k\)-\(s_{k+10}\) plane (top) and as a function of the time step \(k\) (bottom). 
    (b)-(e) Performance of the QRC and the ESN for the predictions of (b) cosine wave, (c) classical MG time-series, (d) integrable quantum spin dynamics, and (e) nonintegrable quantum spin dynamics. 
    NMSE is averaged for \(128\) random realizations for both the QRC and ESN.
    The QRC with \(8\) qubits is constructed at \(a_{\mathrm{in}}=0.001\) and \(a_{\mathrm{back}}=2.5\), and the ESN is equipped with \(8\), \(50\), and \(100\) nodes. 
    The performance of the ESN represents the lowest NMSE achieved through the optimization of network parameters. }
  \label{fig5}
\end{figure}

\begin{figure*}[htbp]
  \centering
  \includegraphics[width=\hsize]{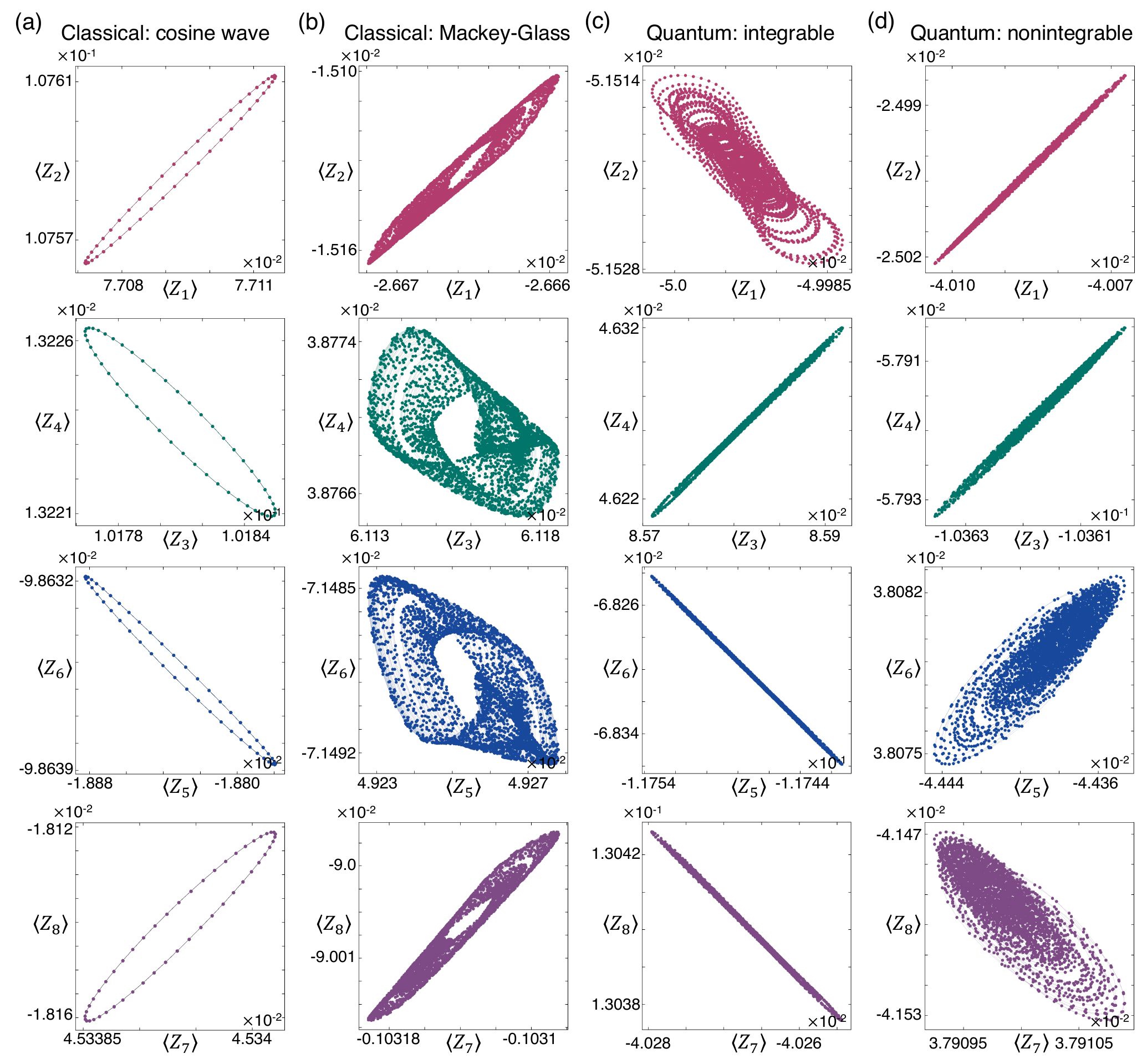}
  \caption{
    Measurement trajectories for a typical \(U_{\mathrm{res}}\) in \(\langle Z_1\rangle\)-\(\langle Z_2\rangle\), \(\langle Z_3\rangle\)-\(\langle Z_4\rangle\), \(\langle Z_5\rangle\)-\(\langle Z_6\rangle\), and \(\langle Z_7\rangle\)-\(\langle Z_8\rangle\) planes. 
    Each point represents \(\left[\langle Z_i\rangle_k,\langle Z_{i+1}\rangle_k\right]\) at cycle \(k\), and each line connects \(\left[\langle Z_i\rangle_k,\langle Z_{i+1}\rangle_k\right]\) and \(\left[\langle Z_i\rangle_{k+1},\langle Z_{i+1}\rangle_{k+1}\right]\). 
    The input and feedback strength are set to \(a_{\mathrm{in}}=0.001\) and \(a_{\mathrm{back}}=2.5\), respectively. 
    The input sequence \(\{s_k\}\) is derived from (a) cosine wave, (b) classical MG time-series, (c) integrable quantum spin dynamics at \((h_x , h_z)= (1.0, 0.0)\), and (d) nonintegrable quantum spin dynamics at \((h_x , h_z)= (1.05, -0.5)\). 
  }
  \label{fig6}
\end{figure*}

In Figs.~\ref{fig5}(b)-\ref{fig5}(e), we present the predictive performance of our QRC across four different time-series: a cosine wave \(\{s_k\}=\{\cos(\omega k )\}\) with \(\omega=\pi/25\), the classical MG time-series with \(t_d=17\), and both the integrable and nonintegrable quantum spin dynamics at \((h_x , h_z)= (1.0, 0.0)\) and \((h_x , h_z)= (1.05, -0.5)\), respectively. 
We employ the input and feedback weights of \(a_{\mathrm{in}}=0.001\) and \(a_{\mathrm{fb}}=2.5\), ensuring that the system remains in the stable phase when subjected to random inputs (Figs.~\ref{fig3} and \ref{fig4}). 
For comparison, we present the performance of an echo state network (ESN), a typical classical reservoir computing framework. 
We optimize network parameters for each \(\tau_f\) while changing the number of nodes; see Appendix \ref{AppendixD} for further details. 

In predicting the cosine wave, the QRC exhibits remarkable performance, surpassing or comparable to the \(100\)-node ESN; \(\mathrm{NMSE}\simeq10^{-9}\) is achieved across all \(\tau_f\) [Fig.~\ref{fig5}(b)]. 
Moreover, it demonstrates pronounced predictive accuracy in forecasting both integrable and nonintegrable quantum spin dynamics, outperforming the \(100\)-node ESN up to  \(\tau_f\simeq 15\) [Figs.~\ref{fig5}(d) and \ref{fig5}(e)]. 
(For \(\tau_f > 15\), both the QRC and the ESN perform poorly with \(\mathrm{NMSE}>10^{-3}\), making comparisons in this regime less meaningful.) 
In contrast, for predictions of the classical chaotic MG dynamics, the effectiveness of our QRC diminishes, falling below the performance of both the \(50\)-node and \(100\)-node ESN for \(\tau_f>3\) as illustrated in Fig.~\ref{fig5}(c). 
These results highlight the strengths and limitations of our QRC in predicting diverse categories of time-series data: it shows a distinct advantage in predicting time-series originating from quantum spin dynamics, yet is less adept at predicting classical MG dynamics.

Notably, our eight-qubit QRC consistently outperforms the ESN with an equivalent number of nodes and, for specific targets, even surpasses the \(100\)-node ESN. 
Corresponding to the number of qubits, our system necessitates the optimization of only \(8+1\) parameters at the output part (including the bias term), which is a significant reduction compared to the \(100+1\) optimization parameters in the \(100\)-node ESN. 
The remarkable performance of the QRC despite having fewer computational nodes can be attributed to the high-dimensionality of the quantum reservoir's Hilbert space, widely recognized as a highly advantageous characteristic for effective reservoir computing~\cite{Fujii:PRA:2017,Tanaka:NeuralNetw:2019}. 
The reduction in computational nodes also contribute to the superior performance of the QRC at \(\tau_f=0\); in such scenarios, where the objective is simply to reproduce the most recent input, excessive complexity can be detrimental. 

In Appendix \ref{AppendixE}, we present the predictive performance of the QRC across various system sizes, utilizing the input signal derived from the nonintegrable quantum spin dynamics. 
Our proposed protocol can be seamlessly extended to different system sizes with analogous circuit architectures as depicted in Fig.~\ref{fig2}(a).
Notably, our QRC demonstrates superior performance compared to the \(100\)-node ESN even when utilizing only \(N=6\) qubits, further underscoring the effectiveness of our approach in processing this quantum-origin input signal. 
In addition, the observed monotonic improvement in performance with increasing system size provides compelling evidence for the potential of further enhancements in larger systems. 

Figure \ref{fig6} illustrates the measurement trajectories under the influence of each time-series signal \(\{s_k\}\); for a typical realization of \(U_{\mathrm{res}}\), \(\left[\langle Z_1\rangle_k,\langle Z_2\rangle_k\right]\), \(\left[\langle Z_3\rangle_k,\langle Z_4\rangle_k\right]\), \(\left[\langle Z_5\rangle_k,\langle Z_6\rangle_k\right]\), and \(\left[\langle Z_7\rangle_k,\langle Z_8\rangle_k\right]\) are depicted. 
When subjected to the cosine wave input, the dynamics of the quantum reservoir system exhibits periodic behavior, manifested as elliptical orbital trajectories in the feature space [Fig.~\ref{fig6}(a)]. 
These trajectories possess a periodicity of \(50\) cycles, which is consistent with the period of the original cosine wave input. 
This observation indicates that the internal dynamics of the quantum reservoir adequately reflects the characteristics of the provided input dynamics. 

In Fig.~\ref{fig6}(b), we present the measurement outcomes influenced by the MG dynamics input. 
These trajectories, presumably reflecting the chaotic attractor observed in Fig.~\ref{fig5}(a), form a complicated manifold that cannot be easily reduced to a simple two-dimensional representation. 
Conversely, under the input signals derived from integrable [Fig.~\ref{fig6}(c)] and nonintegrable [Fig.~\ref{fig6}(d)] quantum spin dynamics, the trajectories collapse into single lines in specific two-dimensional planes: the \(\langle Z_3\rangle\)-\(\langle Z_4\rangle\), \(\langle Z_5\rangle\)-\(\langle Z_6\rangle\), and \(\langle Z_7\rangle\)-\(\langle Z_8\rangle\) planes in Fig.~\ref{fig6}(c), and the \(\langle Z_1\rangle\)-\(\langle Z_2\rangle\) and \(\langle Z_3\rangle\)-\(\langle Z_4\rangle\) planes in Fig.~\ref{fig6}(d). 
Even when trajectories display broader dispersions, they predominantly manifest as superpositions of elliptical curves, each with varying radii and central locations [the \(\langle Z_1\rangle\)-\(\langle Z_2\rangle\) plane in Fig.~\ref{fig6}(c), and the \(\langle Z_5\rangle\)-\(\langle Z_6\rangle\) and \(\langle Z_7\rangle\)-\(\langle Z_8\rangle\) planes in Fig.~\ref{fig6}(d)]. 
This rotation-type reservoir dynamics is attributed to the properties of the provided signals, which originate from quantum spin dynamics that fundamentally involves spin rotations. 

Although the raw time-series signals presented in Fig.~\ref{fig5}(a) alone do not reveal their suitability for the QRC-based prediction, trajectories associated with successful predictions (i.e., cosine and quantum spin dynamics) lead to the inference that input signals inducing rotation-like internal reservoir dynamics are suitable for prediction by the QRC. 
Interestingly, even among such signals, our QRC exhibits varying predictive performance for different time-series. 
For instance, its prediction performance for the nonintegrable quantum spin dynamics is somewhat diminished compared to that for the integrable dynamics, as depicted in Figs.~\ref{fig5}(d) and \ref{fig5}(e). 
This discrepancy might be attributable to the increased complexity of the nonintegrable dynamics, corroborated by the observation that measurement trajectories in the integrable case often converge towards simpler patterns compared to the nonintegrable case. 
Further classification of time-series signals based on their predictability within the QRC framework will be a subject of future research. 

\section{Discussion and conclusion}

In summary, we have developed the feedback-driven QRC framework, which is characterized by the incorporation of feedback connections to the measurement outcomes. 
Similar to existing QRC models, our approach utilizes quantum measurements for information extraction from the quantum reservoir, wherein the quantum state is altered at each cycle. 
Nevertheless, we have demonstrated that memories of previous inputs are effectively reincorporated into the reservoir through feedback processes, which endows our QRC system with the fading-memory property. 
Notably, this framework offers distinct advantages over existing QRC models, particularly in the unrestricted access to the quantum state via projective measurements on all qubits, not limited to weak or partial measurements, without increasing time complexity. 
Moreover, we have elucidated that our QRC system manifests three distinct phases based on the feedback strength, as evidenced by the measurement trajectories. 
The memory performance is maximized when the feedback positions the quantum reservoir system in proximity to the boundary between stable and unstable phases, corresponding to the edge of chaos. 
In time-series predictions, we have shown that the QRC demonstrates proficiency in forecasting dynamics that induces rotations within the quantum reservoir; it outperforms the \(100\)-node ESN for specific prediction targets despite its limited number of optimization parameters. 

Remarkably, the high tunability of our QRC would facilitate task-adaptive performance optimization~\cite{Lee:NatMat:2024}. 
Conventional physical reservoir systems possess few accessible hyperparameters, rendering their reservoir properties largely fixed and making task-dependent optimization a general challenge for reservoir computing frameworks. 
Our protocol will overcome this challenge by flexibly controlling the internal dynamics of the quantum reservoir through varying the input and feedback strengths.  
Combined with its online feature, we anticipate further advancements enabling real-time, on-demand performance tuning through adaptive feedback adjustments. 
Notably, feedback mechanisms are prevalent in the field of quantum machine learning, as exemplified by variational quantum algorithms (VQAs)~\cite{Peruzzo:NatComm:2014,Cerezo:NatRevPhys:2021,Endo:JPSJ:2021}. 
In VQAs, parameters for quantum variational circuits are iteratively trained based on feedback of a cost function calculated from measurement outcomes. 
Although the proposed QRC procedure does not involve variational optimizations, it shares common features with VQAs, particularly in the utilization of measurements and their subsequent feedback. 
Our QRC will thus be further enhanced by leveraging insights from VQAs, especially in feedback configurations. 
Moreover, the measurement and feedback process in our QRC can be interpreted as the construction of a classical representation of the pre-measurement quantum state, analogous to the classical shadow~\cite{Huang:NatPhys:2020}, and the subsequent shadow-based control of the dynamics of the reservoir. 
The potential quantum advantage of our protocol might be demonstrated on the basis of the shadow formalism, which presents an intriguing avenue for future research~\cite{Huang:science:2022-1,Huang:science:2022-2}.

Finally, we highlight the versatility of our feedback-driven QRC scheme, applicable to a broad spectrum of quantum reservoir platforms, encompassing both fermionic and bosonic systems~\cite{npj:Ghosh:2019,Ghosh:CommPhys:2021,Govia:PRR:2021,Kalfus:PRR:2022,Llodr:AdvQuantTechnol:2023}. 
The intrinsic physical properties of these platforms will lead to diverse information processing capabilities, particularly in the types of time-series that can be effectively predicted. 
This emphasizes the importance of investigating a wide array of potential quantum reservoir platforms. 
Notably, such extensive exploration will also provide insights into quantum many-body phenomena from unconventional perspectives, leveraging the effectiveness of the QRC to probe the quantum reservoir system itself~\cite{Kobayashi:arXiv:2023,Kobayashi:arXiv:2024}. 
We anticipate that these investigations will catalyze advancements in both quantum information and quantum materials, thereby propelling the field of quantum information science forward.

\begin{acknowledgments}
K.K. thanks Yukitoshi Motome for insightful discussions. 
The authors acknowledge the support from METI and IPA through the MITOU Target program. 
K.K. is supported by the Program for Leading Graduate Schools (MERIT-WINGS), JSPS KAKENHI Grant Number JP24KJ0872, and JST BOOST Grant Number JPMJBS2418. 
K.F. is supported by MEXT Quantum Leap Flagship Program (MEXT Q-LEAP) Grant No. JPMXS0120319794, JST COI- NEXT Grant No. JPMJPF2014. 
N.Y. is supported by MEXT Quantum Leap Flagship Program (MEXT Q-LEAP) Grants No. JPMXS0118067285 and No. JPMXS0120319794.
\end{acknowledgments}

\appendix

\section{Hardware-efficient reservoir gate}\label{AppendixA}

In step (iii) of our feedback-driven QRC protocol, an eight-qubit reservoir gate \(U_{\mathrm{res}}\) is applied to the entire system to generate a quantum state that encodes both input and feedback information. 
For our numerical calculations with a small system size, we utilize a Haar random matrix for \(U_{\mathrm{res}}\). 
However, such global gate operations are not favorable for practical implementations on physical quantum devices. 
Indeed, the approximation of a Haar-random unitary operation is typically considered to necessitate a circuit with substantial depth (although, realizations with shallow quantum circuits have been recently proposed~\cite{Schuster:arXiv:2024}).

\begin{figure}[t!]
  \centering
  \includegraphics[width=\hsize]{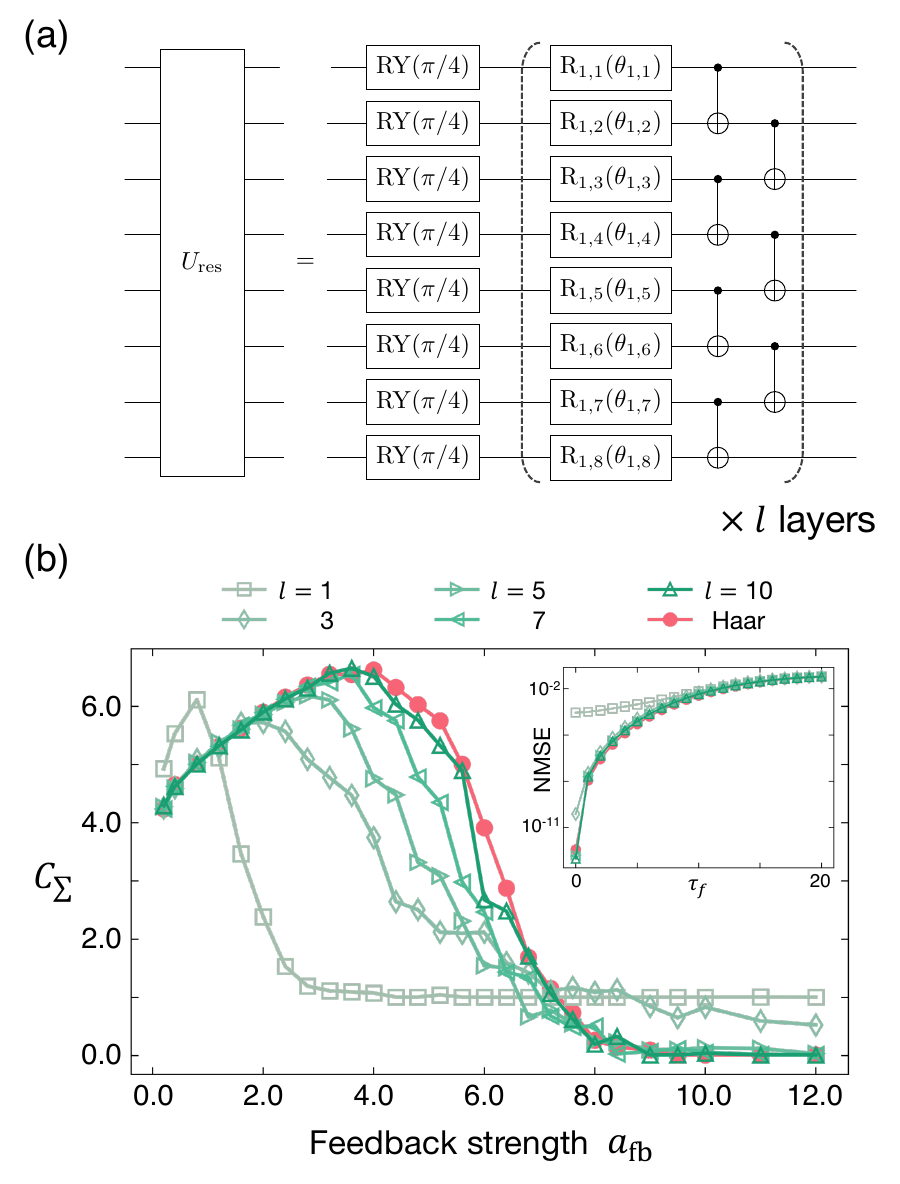}
  \caption{
    (a) Schematic representation of the hardware-efficient circuit for the reservoir gate \(U_{\mathrm{res}}\). 
    The circuit initiates with the application of \(\mathrm{RY}(\pi/4)\) rotations to all qubits, followed by a sequence of \(l\) layers, each comprising randomly generated Pauli rotations and controlled-\(\mathrm{NOT}\) gates. 
    (b) The \(l\) dependence of the total capacity \(C_{\mathrm{\Sigma}}\) in the short-term memory task, and (inset) the corresponding dependence of the predictive performance for the nonintegrable quantum spin time-series. 
    The green lines illustrate the results with varying numbers of layers, and the pink line represents the performance using the Haar random \(U_{\mathrm{res}}\). 
    The input weight is fixed at \(a_{\mathrm{in}} = 0.001\) in both tasks, and the feedback strength \(a_{\mathrm{fb}}\) in the prediction task is optimized for each \(l\): \(a_{\mathrm{fb}} = 0.8\) for \(l=1\), \(3\), \(a_{\mathrm{fb}} = 1.6\) for \(l=5\), \(7\), \(10\), and  \(a_{\mathrm{fb}} = 2.5\) for the Haar random case. 
    Both \(C_{\mathrm{\Sigma}}\) and NMSE are averaged over \(128\) realizations of \(\mathrm{R}_{i,j}\) and \(\theta_{i,j}\).
  }
  
  \label{Sfig1}
\end{figure}

Here, we adopt a hardware-efficient formulation of \(U_{\mathrm{res}}\) comprising only single qubit operations and two-qubit gates among adjacent qubits, originally introduced as ans\"atze for variational quantum eigensolvers~\cite{McClean:NatCommun:2018,Liu:npjqi:2022}. 
Figure \ref{Sfig1}(a) depicts the circuit diagram of \(U_{\mathrm{res}}\), which begins with a wall of \(\mathrm{RY}(\pi/4)\) rotations, followed by a total of \(l\) layers of randomly chosen single qubit rotations and a ladder of controlled-\(\mathrm{NOT}\) gates. 
Specifically, at the \(i\)-th layer, single qubit rotation \(\mathrm{R}_{i,j}(\theta_{i,j})\) is applied to the \(j\)-th qubit in parallel, where \(\mathrm{R}_{i,j}\in \{\mathrm{RX}, \mathrm{RY}, \mathrm{RZ}\}\) is selected with uniform probability and \(\theta_{i,j}\in[0,2\pi)\) is also sampled uniformly. 
Subsequently, controlled-\(\mathrm{NOT}\) gates are alternately applied between adjacent pairs of qubits to efficiently generate entanglement. 
As the number of layers \(l\) increases, the hardware-efficient ans\"atze is expected to converge to the Haar measure, functioning as an efficient scrambler~\cite{Liu:npjqi:2022}. 
We note that \(U_{\mathrm{res}}\), and therefore \(\mathrm{R}_{i,j}(\theta_{i,j})\), remains fixed throughout a single reservoir run encompassing the warming up, training, and testing processes. 

We illustrate the dependence of reservoir performance on the number of entangling layers \(l\) for the short-term memory task in Fig.~\ref{Sfig1}(b). 
In addition, the inset of Fig.~\ref{Sfig1}(b) presents the \(l\)-dependence of the predictive performance for the time-series of nonintegrable quantum spin dynamics. 
When the depth of the entangling layers is insufficient (\(l=1\)), the performance degrades in comparison to those achieved using the Haar random \(U_{\mathrm{res}}\), particularly in the prediction task. 
However, as \(l\) increases, the reservoir performance exhibits improvements in both benchmark tasks, converging rapidly to the value observed in the Haar random case, even with as few as \(l=7\) or \(10\) layers. 
Thus, without the need for the Haar random matrices, our protocol can process information effectively with relatively shallow circuits based on the hardware-efficient implementation of \(U_{\mathrm{res}}\).

\section{Multi-layered feedback connections}\label{AppendixB}

\begin{figure}[t!]
  \centering
  \includegraphics[width=\hsize]{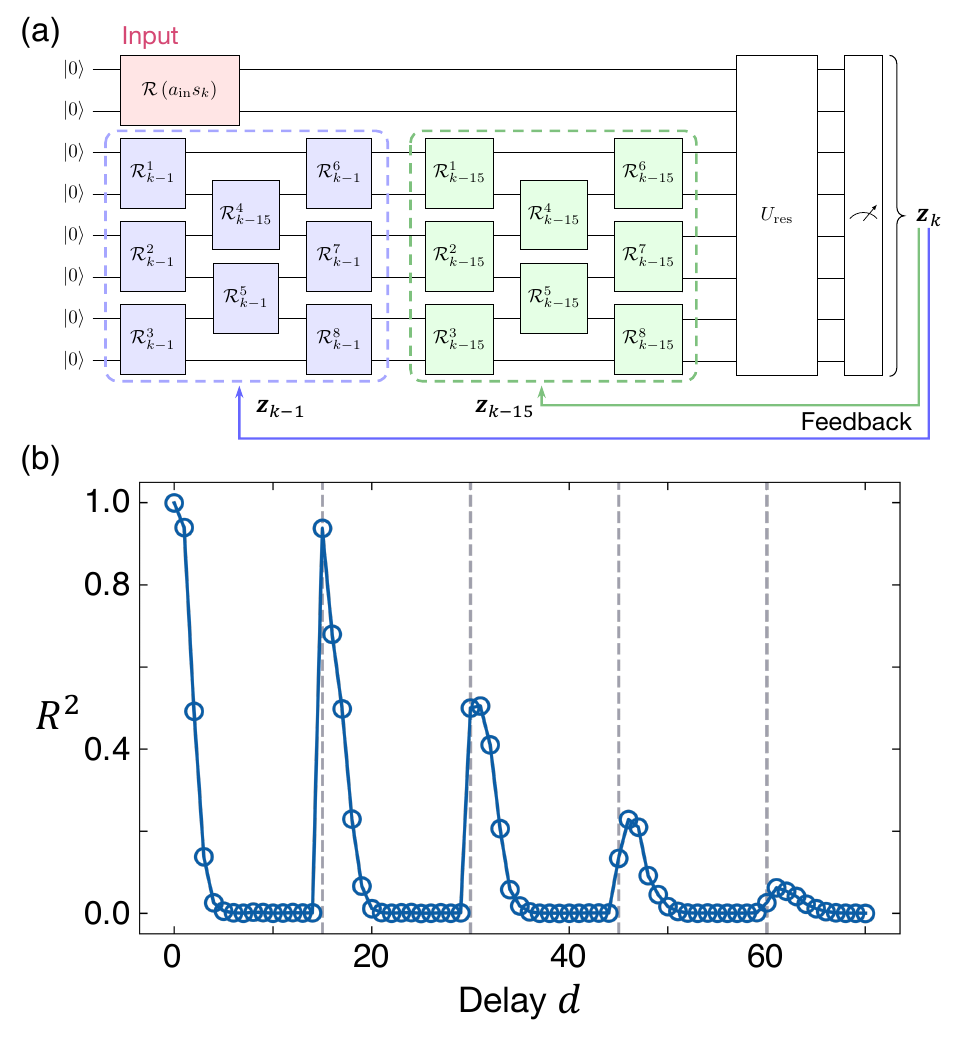}
  \caption{
    (a) Circuit diagram of the QRC architecture at the \(k\)-th cycle, featuring the two feedback layers. 
    The red background denotes the gates for the input, where the randomly sampled input \(s_k \in [0,1]\) is introduced.  
    The blue and green background colors indicate the feedback from the \((k-1)\)-th cycle and the \((k-D_{\mathrm{fb}})\)-th cycle, respectively; we set \(D_{\mathrm{fb}}=15\). 
    The two-qubit operation in Fig.~\ref{fig2}(b) for the \({\alpha}\)-th element of \(\bm{z}_{k-j}\) is concisely represented as \(\mathcal{R}_{k-j}^{\alpha}\equiv\mathcal{R}(a_{\mathrm{fb}}z_{k-j}^{\alpha})\). 
    (b) Reservoir performance for the short-term memory task \(R^2_{d}\) plotted as a function of the delay \(d\). 
    The gray dashed lines represent multiples of \(D_{\mathrm{fb}}\). 
    The performance is averaged over \(128\) instances of \(U_{\mathrm{res}}\), and the input and the feedback strengths are set to \(a_{\mathrm{in}} = 0.001\) and \(a_{\mathrm{fb}}=2.0\), respectively.
    } 
  \label{Sfig2}
\end{figure}

In our QRC protocol, the quantum state at each \(k\)-th cycle is determined by the current input \(s_k\) and the measurement outcomes from the preceding step \(\bm{z}_{k-1}\), with the latter being incorporated through the feedback connections [Fig.~\ref{fig2}(a)]. 
This architecture enables the simultaneous processing of past and present information, which is essential for temporal processing tasks. 
Our feedback-driven architecture is further extensible: the inclusion of additional feedback layers from earlier time steps enhances its capability to process signals with long-term correlations. 
Figure \ref{Sfig2}(a) illustrates the circuit diagram of the proposed feedback-driven QRC at the cycle \(k\), now augmented with additional feedback gates for the measurement outcomes obtained at the (\(k-D_\mathrm{fb}\))-th cycle; we set \(D_\mathrm{fb}=15\). 
The gate configuration for integrating \(\bm{z}_{k-D_\mathrm{fb}}\) is is defined analogously to that employed for \(\bm{z}_{k-1}\) in Fig.~\ref{fig2}(a). 

In Fig.~\ref{Sfig2}(b), the reservoir performance \(R^2_d\) in the short-term memory task for a randomly sampled input sequence \(\{s_k\}\) is depicted as a function of delay \(d\); the feedback strength is set to \(a_{\mathrm{fb}}=2.0\). 
Similar to the results obtained with a single feedback layer [Fig.~\ref{fig3}(a)], the recent input information is effectively encoded within the quantum reservoir system and progressively diminishes upon processing subsequent inputs, as demonstrated by the gradual decline in \(R^2_d\) for small \(d\). 
This is the manifestation of the fading-memory property, which essentially defines the intrinsic timescale as the reservoir computing system. 
Remarkably, despite the almost complete loss of input memory from \(14\) cycles prior (\(R^2_{d=14}\simeq 0\)), memory from \(D_{\mathrm{fb}}=15\) cycles prior is clearly recovered, achieving \(R^2_{d=15}\sim 0.95\). 
Subsequently, a similar progressive decline in memory retention is observed, characterized by a tail of nonzero \(R^2_d\) extending up to \(d \simeq 20\). 
Such an abrupt peak and gradual decline structure also emerges around \(d=30\), \(45\), and \(60\), introducing a timescale of \(D_{\mathrm{fb}}=15\) as all of these are multiples of \(15\). 
This indicates an additional memory mechanism activated by the incorporated feedback layer, facilitating the analysis of input signals with long-term correlations at the corresponding timescale. 
Memories of goal-related information (i.e. \(s_k\), \(s_{k-D_{\mathrm{fb}}}\), \(s_{k-2D_{\mathrm{fb}}},\dots\)) can be concurrently retained and processed within the quantum reservoir. 
We note that this external timescale can be readily adjusted by setting different feedback delay \(D_\mathrm{fb}\), and incorporating multiple feedback layers with different \(D_\mathrm{fb}\) would allow for the capturing of multiple timescales. 
It is worth highlighting that this functionality closely resembles the role of working memory, which transiently hold and manipulate goal-relevant information~\cite{Durstewitz:NatNeurosci:2000,Cucchi:NeuromorphComputEng:2022}. 
As it conflicts with the fading-memory property or echo state property, its implementation in standard reservoir computing frameworks typically requires supplementary structures~\cite{Pascanu:NeuralNetw:2011,Strock:IJCNN:2018,Cucchi:NeuromorphComputEng:2022}. 
In our proposed QRC protocol, the additional feedback layer virtually functions as working memory units, enabling the handling of time-series data with diverse timescales.

\section{Impact of statistical uncertainty in measurements}\label{AppendixC}

In practice, the expectation values of quantum observables are estimated by averaging over a large ensemble of measured values, which inherently introduces statistical uncertainty. 
In our setup, Pauli \(Z\) measurements are performed \(N_{\mathrm{meas}}\) times at each input cycle, and the average of these measured values \(\langle Z\rangle_{N_\mathrm{meas}}\) serves as an approximation to the ideal expectation value \(\langle Z\rangle_{\infty}\). 
However, the statistical uncertainty, which scales as \(O(N_\mathrm{meas}^{-1/2})\), constraints the accuracy of this approximation. 
These statistical fluctuations in measurements not only influence the calculation of the final output \(y_k\) utilizing \(\bm{z}_k\), but also perturb the dynamics of the QRC system itself, as the statistical noise is concurrently fed back along with \(\bm{z}_{k-1}\). 
To evaluate the impact of the statistical error, we simulate the sampling process by introducing Gaussian noise to \(\langle Z\rangle_{\infty}\) with a standard deviation \(\sigma = \sqrt{({1-\langle Z\rangle^2_\infty})/{N_{\mathrm{meas}}}}\)~\cite{Dudas:npj:2023,Mujal:npjqi:2023}.

\begin{figure}[t!]
  \centering
  \includegraphics[width=\hsize]{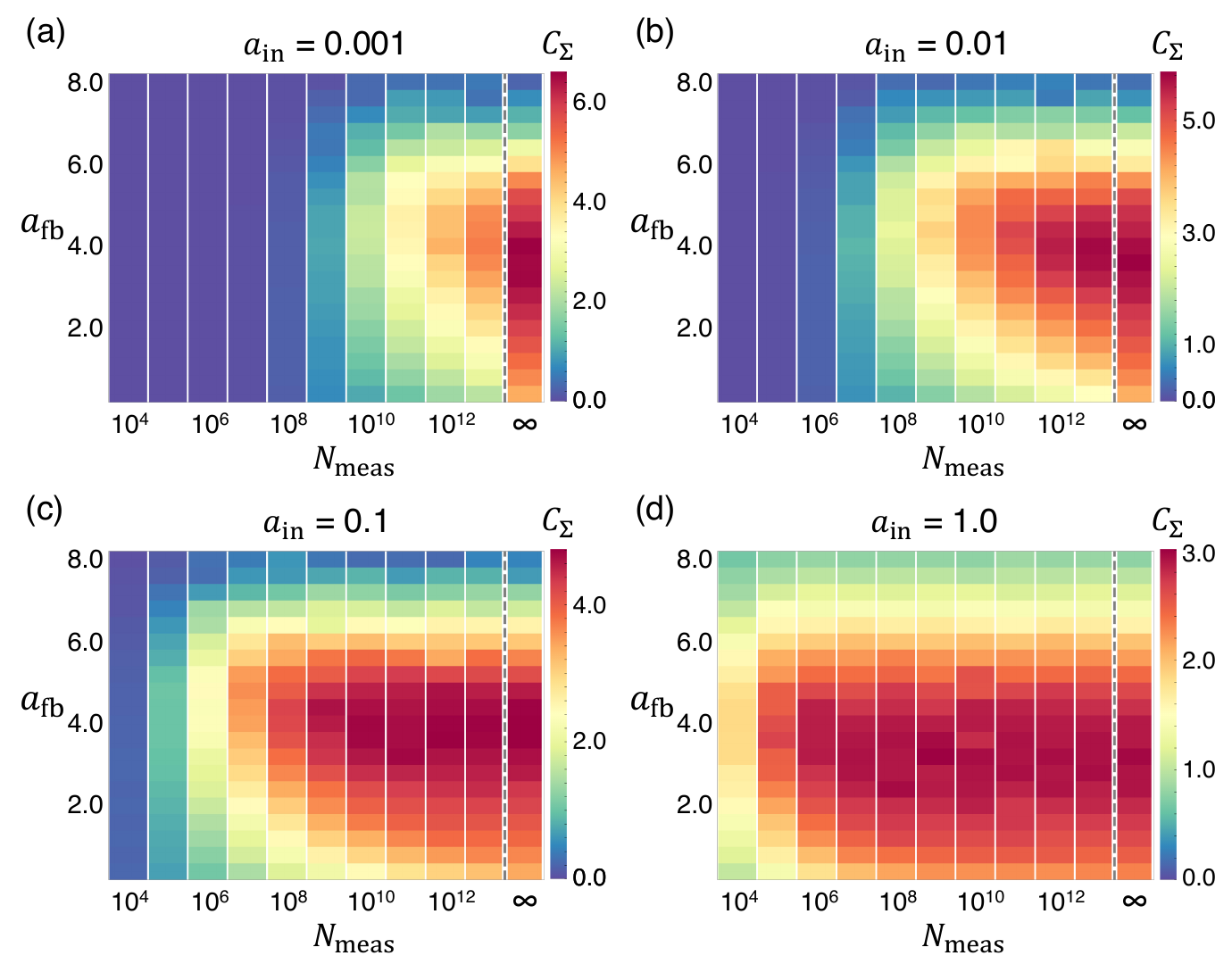}
  \caption{
    The total capacity \(C_{\Sigma}\) as a function of the number of measurements \(N_{\mathrm{meas}}\) and the feedback strength \(a_{\mathrm{fb}}\). 
    The ideal capacities for \(N_{\mathrm{meas}}\rightarrow\infty\) are also shown for reference. 
    The performance is averaged over \(128\) realizations of \(U_{\mathrm{res}}\). 
    The input weight is set to (a) \(a_{\mathrm{in}}=0.001\), (b) \(a_{\mathrm{in}}=0.01\), (c) \(a_{\mathrm{in}}=0.1\), and (d) \(a_{\mathrm{in}}=1.0\).
    } 
  \label{Sfig3} 
\end{figure}

Figure \ref{Sfig3} depicts the total capacity \(C_{\Sigma}\) for the short-term memory task in relation to the number of measurements \(N_{\mathrm{meas}}\) and the feedback strength \(a_{\mathrm{fb}}\) across four different input weights \(a_{\mathrm{in}}\). 
The performance without Gaussian noise is shown for reference [the vertical cut at \(N_{\mathrm{meas}}\rightarrow\infty\) in Fig.~\ref{Sfig3}(a) corresponds to the performance shown in Fig.~\ref{fig3}(b)]. 
Notably, the dependency of the reservoir performance on \(N_{\mathrm{meas}}\) is significantly influenced by the input weight \(a_{\mathrm{in}}\). 
For a small input weight (\(a_{\mathrm{in}}=0.001\)), \(C_{\Sigma}\) remains almost zero irrespective of \(a_{\mathrm{fb}}\) up to \(N_{\mathrm{meas}}=10^8\), due to the disturbance caused by the statistical noise [Fig.~\ref{Sfig3}(a)]. 
In other words, when \(N_{\mathrm{meas}}\) is insufficiently large, statistical fluctuations dominate over the impacts of the input operation, leading to inadequate information encoding. 
Nevertheless, this issue can be addressed either by reducing statistical noise through an increased number of measurements \(N_{\mathrm{meas}}\), or by amplifying the input effect via a larger input weight \(a_{\mathrm{in}}\). 
Indeed, as \(N_{\mathrm{meas}}\) increases further, the total capacity \(C_{\Sigma}\) gradually becomes nonzero, recovering the three phases based on the feedback strength \(a_{\mathrm{fb}}\) observed in Sec.~\ref{sec3}.

Moreover, as the input weight increases to \(a_{\mathrm{in}} = 0.01\) and \(0.1\) [Figs.~\ref{Sfig3}(b) and \ref{Sfig3}(c)], the region of nonzero \(C_{\Sigma}\) expands towards smaller values of \(N_{\mathrm{meas}}\). 
For even larger input weight (\(a_{\mathrm{in}} = 1.0\)), the total capacity rapidly converges to that achieved with ideal expectation values around \(N_{\mathrm{meas}}\simeq 10^5\) [Fig.~\ref{Sfig3}(d)]. 
Additionally, a strong feedback \(a_{\mathrm{fb}}\) also contributes to the robust operation of the QRC system, as it similarly prevents the dominance of statistical fluctuations. 
This is evidenced by the faster decay of \(C_{\Sigma}\) with decreasing \(N_{\mathrm{meas}}\) in the small \(a_{\mathrm{fb}}\) region compared to the large \(a_{\mathrm{fb}}\) region, as clearly shown in Fig.~\ref{Sfig3}(b). 
These observations indicate that the reservoir performance in the presence of statistical fluctuations is governed by a balance among the three factors: input, feedback, and statistical uncertainty. 
Hence, although statistical fluctuations introduce some disturbances into the dynamics of the QRC, their impact can be mitigated by flexibly adjusting the input weight \(a_{\mathrm{in}}\) and the feedback strength \(a_{\mathrm{fb}}\). 

It is noteworthy that the repetition of measurements does not compromise the online nature of the feedback-driven QRC in Fig.~\ref{fig1}(d). 
To obtain multiple ensembles \(m = 1,\dots N_{\mathrm{meas}}\) for averaging at the \(k\)-th input cycle, it suffices to independently repeat the dynamics at the \(k\)-th cycle \(N_{\mathrm{meas}}\) times, thereby preserving online operation without the need to revisit earlier input cycles. 
In contrast, in the continuous monitoring QRC protocols with indirect measurements in Fig.~\ref{fig1}(c), after a single reservoir run (\(m=1\)) up to the \(L\)-th input cycle, the measurement results for every \(k\)-th input cycles \([Z_{k=1}^{m=1},Z_{k=2}^{m=1}\dots,Z_{k=L}^{m=1}]\) are simultaneously obtained. 
Upon the subsequent reservoir run (\(m=2\)), the QRC system restarts from the beginning, yielding \([Z_{k=1}^{m=2},Z_{k=2}^{m=2}\dots,Z_{k=L}^{m=2}]\); this process is repeated \(N_{\mathrm{meas}}\) times to calculate the average \([\langle Z_{k=1}\rangle_{N_{\mathrm{meas}}},\dots,\langle Z_{k=L}\rangle_{N_{\mathrm{meas}}}]\). 
Therefore, to facilitate online operation in the continuous monitoring QRC protocols [Fig.~\ref{fig1}(c)], parallelization with independent \(N_{\mathrm{meas}}\) threads, each concurrently generating the \(m\)-th outcome for averaging (as demonstrated in quantum optical platforms~\cite{Garcia-Beni:PRAppl:2023}), or measurement setups capable of obtaining expectation values in a single run (such as NMR), would be required. 
Otherwise, after processing up to the final input \(s_{k=L}\) in the \(m\)-th iteration, it would be necessary to return to the initial input \(s_{k=1}\) to initiate the \((m+1)\)-th measurement iteration. 
This virtually undermines the online nature of the system, as it necessitates referencing past inputs even after the current input has been provided.

\section{Echo state network (ESN)}\label{AppendixD}

ESN is a well-established framework in classical reservoir computing~\cite{Jaeger:GMD:2001,Jaeger:Science:2004,Jaeger:NeuralNetw:2007}. 
For an ESN characterized by its number of nodes \(N_{\mathrm{node}}\), the internal state at cycle \(k\) is represented as an \(N_{\mathrm{node}}\)-dimensional vector \(\bm{z}^{\mathrm{ESN}}_{k}\). 
This state is updated iteratively by
\begin{equation}
  \bm{z}^{\mathrm{ESN}}_{k+1} = (1-\mathrm{lr})\bm{z}^{\mathrm{ESN}}_{k} + \mathrm{lr}f(W_\mathrm{in}s_{k+1}+W\bm{z}^{\mathrm{ESN}}_{k}),
\end{equation}
where \(\mathrm{lr}\) signifies the learning rate, and \(f\) represents the element-wise activation function, for which we adopt \(f(x)=\tanh(x)\). 
\(W_{\mathrm{in}}\) represents the \(N_{\mathrm{node}}\)-dimensional input weights, each assigned values of either \(1\) or \(-1\) with an equal probability. 
\(W\) denotes the \((N_{\mathrm{node}}\times N_{\mathrm{node}})\)-dimensional recurrent weights matrix, whose elements are sampled from a Gaussian distribution and scaled to achieve a spectral radius of \(\lambda\). 
These internal weights remain fixed throughout the process. 
In the learning phase, the training matrix \(X\) is constructed using the state vector \({\bm{z}^{\mathrm{ESN}}_{k}}'=[{\bm{z}^{\mathrm{ESN}}_{k}}^\top,1]^\top\) in the same manner as Eq.~(\ref{eq2}). 
The output weight \(\bm{w}\) is trained via Ridge regression with a regularization term \(\beta=10^{-5}\) to prevent instability, expressed by
\begin{equation}
  \bm{w} = (X^\top X+\beta I)^{-1}X^\top\bar{\bm{y}}_{\mathrm{tr}}.
\end{equation}
The final output at the \(k\)-th cycle is computed as \(y_k = {\bm{z}^{\mathrm{ESN}}_{k}}'^\top \bm{w}\), and the performance is evaluated using Eq.~(\ref{eq5}). 
In Figs.~\ref{fig5}(b)-\ref{fig5}(e), we present the optimal performance of the ESN by varying the parameters within the ranges \(\lambda\in [0.5,0.75,0.95,1.25,1.5]\) and \(\mathrm{lr}\in [0.4,0.6,0.8,1.0]\). 

\section{Feedback-driven QRC with varying system size}\label{AppendixE}

\begin{figure}[t!]
  \centering
  \includegraphics[width=\hsize]{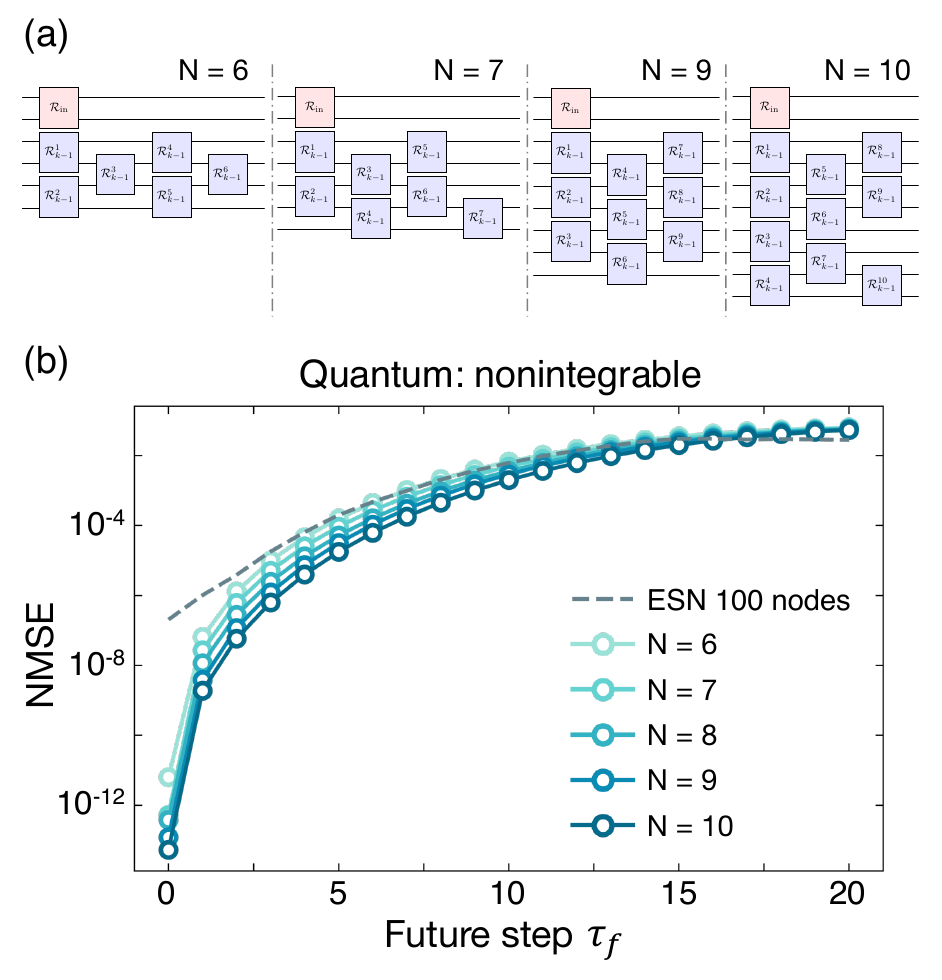}
  \caption{
    (a) The gate configurations for the input and the feedback parts with system sizes \(N=6\), \(7\), \(9\), and \(10\). 
    The red and blue background colors represent the gates for the input \(\left[\mathcal{R}_{\mathrm{in}}=\mathcal{R}(a_{\mathrm{in}}s_k)\right]\) and feedback \(\left[\mathcal{R}_{k-1}^i=\mathcal{R}(a_{\mathrm{fb}}z_{k-1}^i)\right]\), respectively. 
    The initial state \(|\psi\rangle_0=|0\rangle^{\otimes N}\), the \(N\)-qubit Haar random reservoir gate \(U_{\mathrm{res}}\), and the measurement on all qubits are consistent with the \(N=8\) case in Fig.~\ref{fig2}(a), omitted here for simplicity. 
    (b) Performance of the QRC in predicting signals from the nonintegrable quantum spin dynamics. 
    The input weight is fixed at \(a_{\mathrm{in}} = 0.001\), and the feedback strength \(a_{\mathrm{fb}}\) is optimized for each \(N\): \(a_{\mathrm{fb}} = [1.5,2.0,2.5,3.5,5.5]\) for \(N=[6,7,8,9,10]\), respectively.  
    NMSE values are averaged over \(128\) realizations of \(U_{\mathrm{res}}\).
    The performance of the ESN with \(100\) nodes is also represented as a reference, whose network parameters are optimized as detailed in Appendix \ref{AppendixD}.
  }
  \label{Sfig4}
\end{figure}

We investigate the system size dependence of the performance of the feedback-driven QRC, particularly focusing on the prediction task for the nonintegrable quantum spin dynamics. 
Figure \ref{Sfig4}(a) presents the gate configurations for the input and feedback parts across various system sizes \(N=6\), \(7\), \(9\), and \(10\). 
These configurations are defined similarly to the case with \(N=8\) in Fig.~\ref{fig2}(a); however, it is important to note that our architecture is not limited to these specific configurations. 
It can be expanded as necessary, for instance, by integrating additional input gates. 
We also note that our QRC is scalable to accommodate even larger systems without encountering the barren plateau problem~\cite{McClean:NatCommun:2018}, as the classical processing component only involves the linear regression of measurement outcomes. 
The remaining circuit elements, including the initial state \(|\psi\rangle_0=|0\rangle^{\otimes N}\), the \(N\)-qubit Haar random reservoir gate \(U_{\mathrm{res}}\), and the measurement on all qubits, remain consistent with the circuit depicted in Fig.~\ref{fig2}(a).

In Fig.~\ref{Sfig4}(b), the predictive performance is plotted as a function of the future prediction step \(\tau_f\), quantified by NMSE. 
The performance of the \(100\)-node ESN is also presented for comparison. 
In the early time predictions with small \(\tau_f\), NMSE markedly decreases as \(N\) increases. 
For instance, at \(\tau_f=1\), NMSE with \(N=10\) is approximately \(50\) times lower than that with \(N=6\); similar improvements of over \(10\) times are observed for other \(\tau_f\leq 5\). 
The trend of monotonically improved performance with increasing \(N\) persists even at larger \(\tau_f \lesssim 15\). 
These significant enhancements in predictive accuracy can be ascribed to the exponentially expanding Hilbert space as a function of \(N\). 
Since the primary role of the reservoir is to project input information onto a high-dimensional internal space, the increase in the dimensionality of the Hilbert space directly enhances the reservoir performance. 
Indeed, the monotonic decrease in NMSE with increasing \(N\) strongly suggests the potential for further performance improvements with larger system sizes. 
Notably, our QRC outperforms the \(100\)-node ESN even with \(N=6\) qubits, despite having only \(6+1\) optimization parameters. 
This further highlights the efficacy of our QRC in processing the quantum-origin input signal.

\bibliography{bibtex}

\end{document}